\documentclass{article}
\usepackage{graphicx,latexsym}
\usepackage[widemargins]{a4}
\begin{document}

\title{Evolution dynamics in terraced NK landscapes}
\author{Paolo Sibani  and  Andreas Pedersen\\ 
 Fysisk Institut,  SDU-Odense Universitet\\
Campusvej 55, DK5230 Odense M, Denmark}
\maketitle 

\begin{abstract} 
We consider populations of agents  evolving in  
the  fitness landscape of an extended $\bf NK$ model 
with a  tunable amount of neutrality.
We study the statistics of the jumps in 
mean population fitness which occur 
in the `punctuated equilibrium' regime
and show  that, for  
a wide range of landscapes parameters, the  
number of events  in time $t$  is Poisson distributed,
with the time parameter replaced by the logarithm of time.
This simple log-Poisson statistics 
likewise describes the number of records 
in any  sequence of $t$ independently  generated random numbers. 
The implications of such behavior  for evolution dynamics are   discussed. 
\end{abstract}

 
\noindent {\bf Introduction.}
Evolutionary  dynamics  can be described  as a 
stochastic process unfolding in a    
`fitness landscape'\cite{Wright31},
a process  which 
can be  simulated  by means 
of genetic algorithms \cite{Holland75}.  
The   dynamical behavior of  such algorithms 
is  controlled by a number of parameters\cite{vanNimwegen97}
as   e.g. the population size,   the rate of mutation and  
the strenght of selection. 

A  pervasive dynamical regime is 
the so-called    `punctuated equilibrium'\cite{Eldredge72}
or `epochal behavior', where     
relevant measures of  evolution, as e.g. the  mean fitness,  remain 
constants  for long periods  during which 
the  fitness distributions of the individuals 
is strongly peaked. Occasionally,   a fitter mutant appears
and quickly `takes over'   the population (see e.g. Fig.~1). 
Real experiments   performed on bacterial colonies
evolving in a controlled  environment have shown 
that the fitness and cell size increase at
a  decelerating rate \cite{Lenski94}.
A similar slowing down is discussed by  Kauffman\cite{Kauffman95}   
in  the  `long jump' dynamics of the $ \bf NK$ model, while  
Aranson et al.\cite{Aranson97}   find  a logarithmic growth of the
average fitness for quasi-species evolving in a rugged fitness landscape.

In a macroevolutionary context,  Raup and 
Sepkoski\cite{Raup82} suggested that the  
noticeable decay of the extinction rate\cite{Newman98,Sibani95}
might  stem from the  properties of 
an underlying  optimization process.  
This idea was  taken up in the `reset' model\cite{Sibani95,Sibani97},
where   jumps in the average  fitness of populations
are linked to  fitness record achieved
during evolution\cite{Alstrom99}.
Such a link between small and  grand scale evolution
is rather  controversial: If    macroevolutionary    events   
are mainly  driven  extrinsically,
 e.g. by  meteorite impacts\cite{Alvarez80}, any patterns 
in the fossil record   must ultimately  stem from the mechanics of  
celestial bodies  in  the solar system\cite{Raup84}.
Conversely, if, as recently proposed by several authors\cite{Newman99b},
the fossil record is mainly shaped by 
complex  interactions within the  biotic system,   
macro-evolutionary patterns  must   emerge from  population dynamics.
 
Our main interest lies in the  statistical properties  of 
the  evolutionary  jumps  underlying  fitness changes. 
Since the role  of neutral mutations\cite{Kimura83} for evolution
is well established,  and since    punctuated equilibria 
 exist even in the absence of local fitness 
maxima\cite{vanNimwegen97,Bornholt98}, we chose to   
study evolution on  `terraced'  landscapes with 
a tunable degree of neutrality.

We find that, on average,  the number of jumps taking place in time
 $t$ grows proportionate to  $\log t$, and that 
the rate of events consequently 
decays as $\approx 1/t$.  Secondly and most importantly,  we show that  
the record   dynamics  \cite{Sibani95,Alstrom99}  provides a  
reasonable description of  microevolution, with some limitations which
are also outlined.  Thirdly, we emphasize 
that  power-law  decays generically characterize
the  correlation functions  of   time series generated
by a record driven dynamics.

\begin{figure}[t]
\centering
\includegraphics[width=0.7\textwidth]{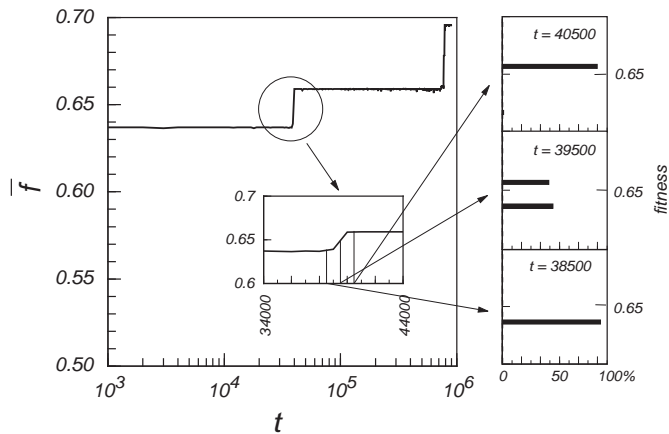}
\vspace{-0.5cm}
\caption{The mean fitness $\overline{f}$ of  a
 single population  is shown in the main panel
as a  function of time. The right panel shows the
 fitness distributions at times $t=38500$, $t=39500$ and $t=40500$.
 The simulation  parameters are $N=64$, $K=31$, $n=1000$,
 $\mu_f=0.5$, $\sigma_f=2.5\times10^{-3}$,
 $\beta=1$, $u=10^{-4}$ and $j=10^9$, corresponding to the original $\bf NK$ model.}
\label{fitness}
\end{figure}

\noindent {\bf Method.}  
 Each `genome' in a population constitutes   a point in an abstract 
configuration space usually called  a  fitness landscape \cite{Wright31}.  
To  construct such a landscape we use an elegant 
  prescription due to Kauffman, the widely
 known $\bf NK$ model\cite{Kauffman87}:
We represent  genomes as  strings 
of ${\bf N}$ bits $ {\bf x} = (x_1,x_2,\ldots x_N)$,  
each being either $0$ or $1$. 
The fitness $F(\bf x)$ of configuration $\bf x$ is  defined as 
\begin{equation}
F({\bf x}) = \frac{1}{\bf N}\sum_{i=1}^{\bf N} f_i,
\label{Kauffman1}
\end{equation}
where the contribution from site $i$,  $f_i$,
 is a random function  
  depending  on $x_i$  and  $\bf K$ other 
 $x_l$'s.
More precisely, $f_i$ is a random function
of  $2^{{\bf K}+1}$   arguments with  values uniformly distributed in  
$(0,1]$. We let $\mu_f$ and $\sigma_f$ be the 
average and spread of the distribution from which
the $f_i$ values are generated.

If $\bf K$ is zero, the change in fitness due to
the change of one   $x_i$ (a point mutation) 
is  of order $1/N$, and the landscape may be 
regarded  as   smooth.
By way of contrast, when  $\bf K = {\bf N}-1$  
a single  point mutation changes all the $f_i$'s,   and  
the landscape becomes   `rugged'. Intermediate cases  
correspond, of course, to intermediate   $\bf K$ values.

Our  version of the $\bf NK$ model is modified in
two respects. Firstly, the sum in 
Eq.\ref{Kauffman1} is  shifted by $\mu_f$ and scaled by $\surd {\bf N} $ rather than $N$.
 This ensures that the distribution of 
fitness values keeps the same variance for any  value of  $N$.
More importantly,  we introduce   
tunable  neutrality by  discretizing 
 the fitness values  into `terraces',   according to the formula: 
\begin{equation}
  F'({\bf x}) = \frac{{\rm nint  }(j \surd {\bf N}  [ F({\bf x}) - \mu_f ] + j \mu_f)}{j}.  
\label{Kauffman2}
\end{equation}
Here  $\rm nint$ stands
for the nearest integer function, and  $j$ denotes 
the number of terraces. For small $j$  there are few broad
 terraces, while for 
 $j \rightarrow \infty$  the fitness  values 
 approach a continuum and the original $\bf NK$ model  is regained.
Accordingly, one expects that, as seen e.g. in Fig.~2, the effect of
varying $j$ should be  stronger for small values of $j$.

In the simulations, a configuration ${\bf x} $ is 
cloned with a probability  
$ p \propto  \exp(\beta F({\bf x}))$.
With probability $u <<1$, the cloned string  undergoes  a one-point
mutation at a  randomly chosen locus. 
Finally, a random and fitness independent    deletion mechanisms
is applied,  which keeps the   population size fluctuating 
around a fixed average $n$.
Both the generation $/$ mutation part  and the
deletion part of the algorithm are performed sequentially.
Subsequently,  the information about the new fitness distribution
is incorporated in the cloning probabilities.  
This entire  process counts as one update and defines the unit of time.

Punctuated equilibrium dynamics requires that a fitter mutant 
be able to survive and   spread  in the population on a time
scale  short compared to the inverse mutation frequency.
Therefore, $u$ should not be too high and $\beta$ should not be
too low. Within these constraints  $u$ should be as high as possible,
in order to have a good statistics within the time window 
of the computation. Also, too high a $\beta$ value quenches
the dynamics completely. These design consideration lead to 
the values of $u$ and $\beta$ used throughout the calculations.  

 A concise  description of the 
dynamics is provided by the average $\overline{f}(t)$ of the
 distribution of fitness values through the population.
  Such a {\em trajectory} is shown
in the main panel of Fig. 1 to consist of a number of flat plateaus 
separated by  rather well defined jumps. The number $m(t)$   of   jumps   
 occurring in the interval $[0,t)$ is a stochastic process
whose distribution    can be sampled  by repeating the 
simulations or, equivalently, by considering an ensemble of
landscapes, where different trajectories are generated by 
 independently updating each system. The
ensemble average and variance of $m$ are denoted
by $E(m)$ and $\sigma^2(m)$ respectively,
while $a_m(t)$ and $v_m(t)$ are the corresponding estimators.
The notation describing the inputs and results 
of our simulations is summarized in Table~1.

\begin{table}[t]
\begin{center}
\begin{tabular}{l|c|r}
\hline
\hline
variable & meaning & values    \\
\hline
\hline
$\bf N$ & genome length & $64$\\
${\bf K}$ & degree of ruggedness & $7, 15$ and $31$\\
$n$ & population size & $1000$ \\
$\sigma_f$ & spread of $f_i$'s  in Eq.\ref{Kauffman1} &  $  2.5 \times 10^{-3} $ \\
$\mu_f$ & mean of $f_i$'s in Eq.\ref{Kauffman1} & $0.5$ \\
$j$ &  number of terraces &  $10$ - $10^9 \approx \infty$ \\
$\beta$ & reproductive selectivity & $1$\\
 $u$  &  mutation prob. per cloning & $10^{-4}- 5 \cdot 10^{-4}$\\
$s$ &  ensemble size & $s=80$ \\ 
$\overline{f}$ & population averaged    fitness &  see Fig.~1 \\
$m$ & no. of evol. steps & see Fig.~2 \\
$a_m(t)$ & sample average of $m$ & - \\
$\lambda$ &   $\lambda \log t \approx a_m(t) $ & see slopes of Figs.~2 and 3\\
\hline
\end{tabular}
\caption{The table summarizes the notation and parameter valuse used in the text.}
\end{center}
\end{table}

\noindent {\bf Results.}
The basic qualitative features of the data are expressed by   Fig.~1:
Its left panel   shows the mean fitness $\overline{f}$ 
of a   single  population on a semilogarithmic scale.
The  right panel details the behavior  close to
the evolutionary jump   at  
$t \approx  3 \times 10^4$ by depicting  the
 distribution of fitness values  right before, 
 during and   right after the jump. Nearly all
strings  have the same fitness  in the initial and
  final situations, while the  transition stage
features two different fitness values. 
Punctuated equilibrium behavior  and a very peaked   
fitness distribution are widely found 
in previous studies\cite{Weisbuch92,Newman98b} as 
well as in our simulations.
\begin{center} 
\begin{figure}
\centering
 \includegraphics[width=0.7\textwidth]{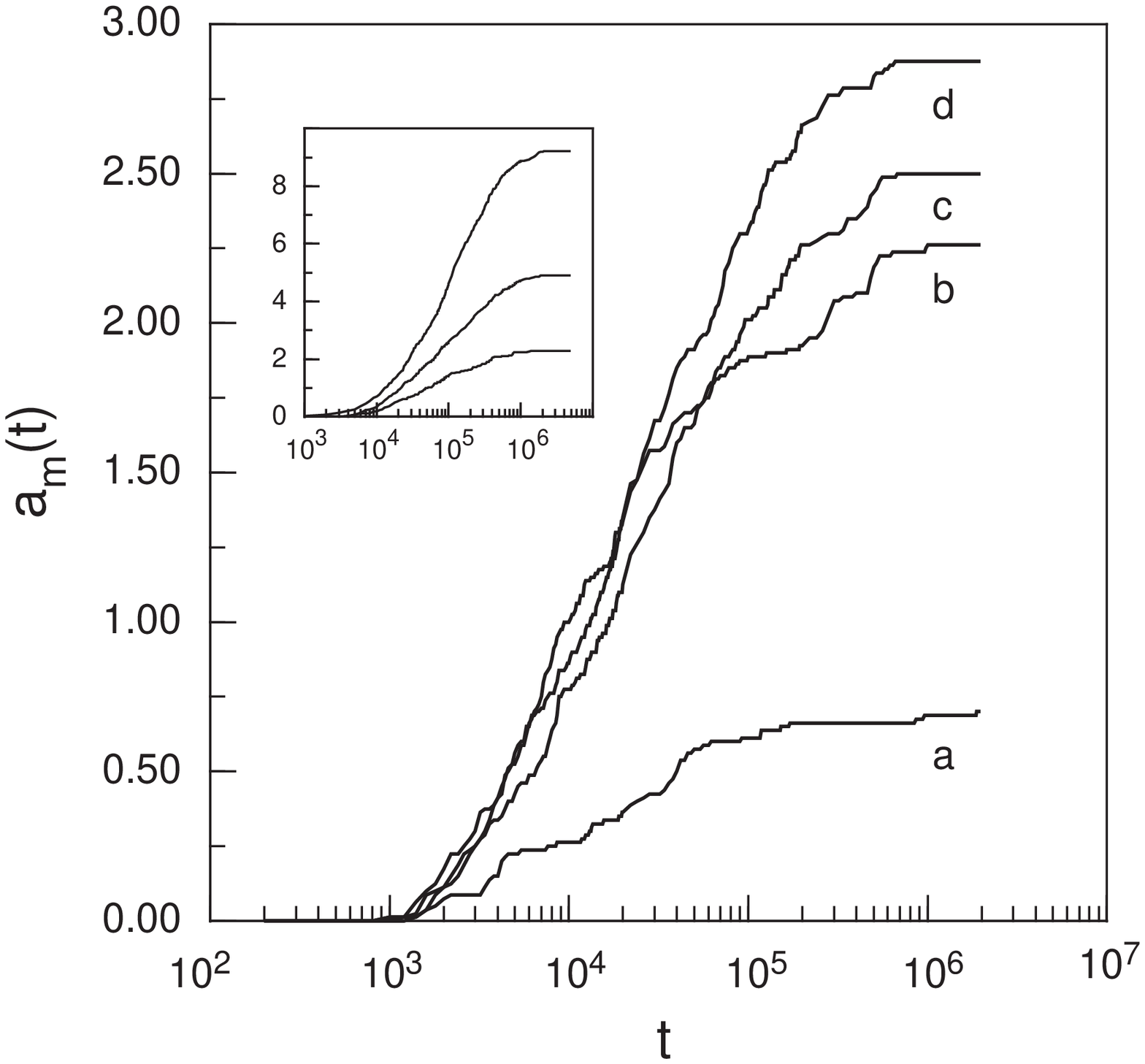}
\vspace{-3cm}
\caption{Sample average of the number of fitness jumps, $a_m$ 
as a function of time $t$ for landscapes
with   ${\bf K}=31$ and   $a: j=10$, $b: j=10^2$, $c: j=10^3$ and $d:j=10^9$.
 The mutation rate is $u=5\times10^{-4}$. 
The insert shows the same quantity, but  
for landscapes with no terraces ($j = 10^9$) and with  ${\bf K}=7$ (top curve), ${\bf K}=15$ and ${\bf K} =31$ (bottom curve).
The mutation rate is here $u=10^{-4}$.}
\label{average}
\includegraphics[width=0.7\textwidth]{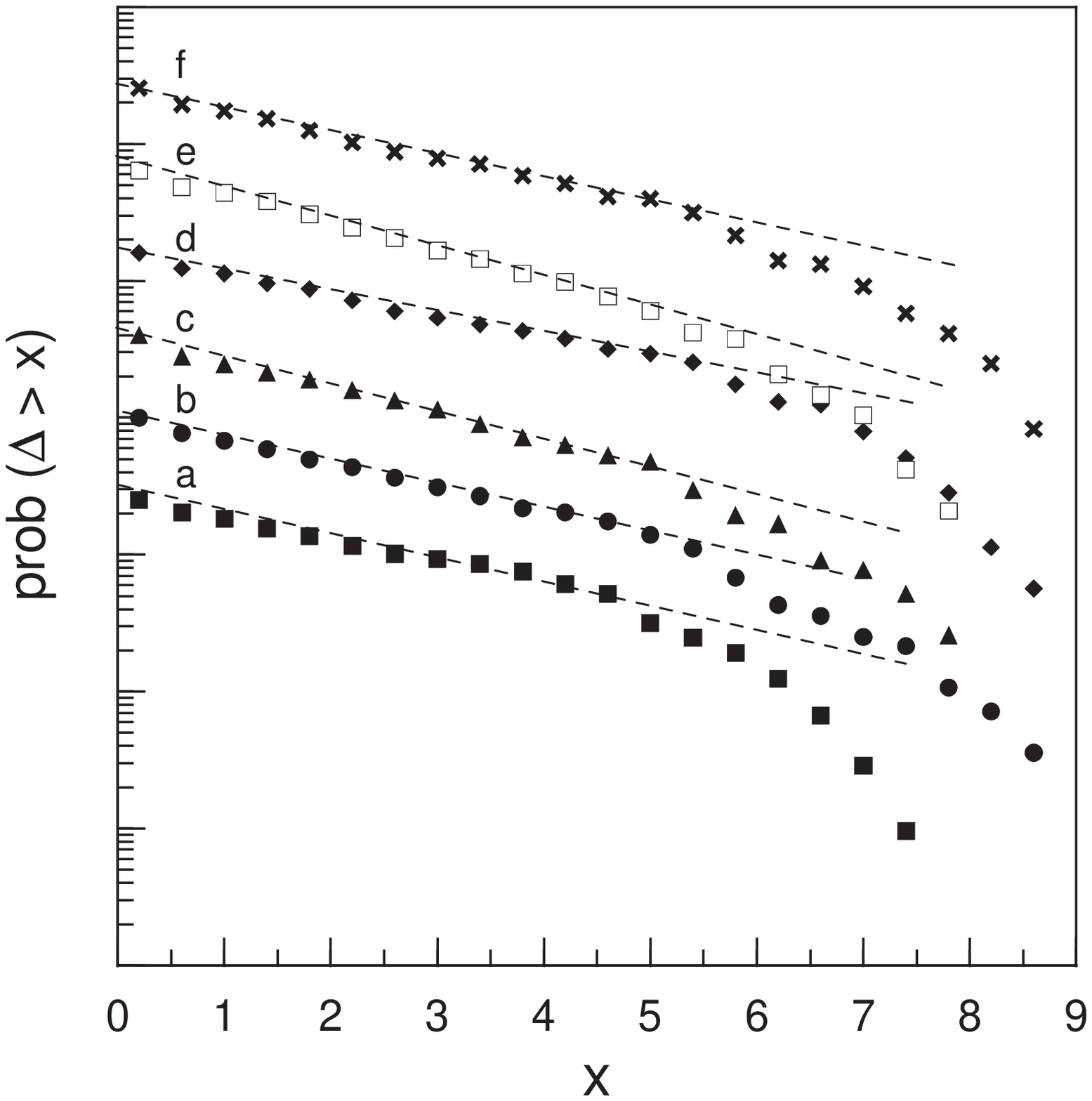}
\vspace{-3cm}
\caption{Distribution of log-waiting times for landscapes with 
${\bf K}=31$ and   $a:j=10^2$, $b:j=10^3$, $c:j=10^4$, $d:j=10^5$, $e:j=10^6$ and $f:j=10^9$.
For graphical clarity the curves are vertically shifted relative to each other 
by a factor of four.  The mutation rate is  $u=5\times10^{-4}$.}
\label{waiting_times}
\end{figure} 
\end{center}
For a more quantitative data analysis, the 
statistical properties of the jumps and their associated 
waiting times must be studied. 
We  let $m_i(t)$ be the number of jumps occurred at time $t$  in  
trajectory  $i$,   and consider the  sample
average:   
\begin{equation}
a_m(t) = \frac{\sum_{i=1}^s m_i(t)}{s},
\end{equation}
and the sample variance:  
\begin{equation}
v_m(t) = \frac{\sum_{i=1}^s (m_i(t) - a_m(t))^2}{s-1}
\end{equation} 
as functions of time.
To calculate   error bars on $a_m$ and $v_m$ we 
need the corresponding  standard deviations, which  are 
$\sigma(m)/\surd s$ and
 $ \surd (\sigma^2(m^2) +8 E^2(m) E(m^2) -4 E(m)(E^3(m)+E(m^3)) )/\surd s + {\cal O}(1/s)$.
The latter relation results from straightforward but rather tedious algebra. To lowest order 
in $ s^{-1/2}$ one may now replace the moments of $m$ with their 
sample estimators. In general,   the relative errors on various quantities
of interest are of order $1/\surd s  \approx 10 \%$.

Fig.~2 shows the average number of jumps, for a number of different 
parameters values, as a function of time.  
$E(m)$ is seen to grow logarithmically,    with 
a strong ${\bf K}$ dependence   slope, for `short'
log-times. The leveling off noticed 
at large times   stems from  the fact  
that, as the fitness increases,
fitness improvements 
become progressively smaller. Eventually, they
  get lost in the noise, rather than triggering a jump.
At this point  record statistics and fitness  evolution  
must part company. 

For large $t$, the probability of   $n$ records 
 in a sequence of $t$ independently
 drawn random numbers is given 
 by\cite{Sibani93}: 
\begin{equation}
P_t(n) = \frac{(\log \lambda t)^n}{n!}  t^{-\lambda}.
\label{logpoisson}
\end{equation}
This is  a Poisson distribution with  $\log t$ replacing the
time argument. The strength parameter  $\lambda$ 
describes the possibility that 
many  searches for records  take  place independently and
in  parallel and/or  the 
situation  where    records remain undetected.
In our systems we observed that several records were indeed lost
in the noise, and did not   trigger  any evolutionary event.
 
A  mathematically equivalent description of 
record dynamics is 
provided by the  distribution of the   variables
\begin{equation}
\Delta_k = \log t_k -\log t_{k-1} = \log(t_k/t_{k-1}).
\label{expo}
\end{equation}
It follows from  standard   arguments that in a (log) Poisson 
process   these  $\Delta_k$ are  independent and
have the common distribution:
\begin{equation}
P(\Delta >x) = \exp(-\lambda x).
\label{waittime}
\end{equation}

The  empirical distribution of the $\Delta_k$'s is shown on a semilogarithmic
scale in Fig.~3  for  ${\bf K} =31$   and for differing degrees
of terracing. The 
decay of  $P(\Delta >x)$ seems quite  well described by an exponential for 
$x \leq 5$, with  the tail of the distribution falling  off more rapidly,
likely due to the inevitably  poor   sampling of large $x$ values in a finite
time simulation. Banning the effect of the deviations from  pure
record statistics, $\lambda$ is,  by Eqs.~\ref{logpoisson} and \ref{waittime},
the slope of $a_m(t)$ vs. $\log t$ as well as  minus the slope of 
$\log P(\Delta >x)$ vs $x$. We calculated these slopes from the data in both ways  (cutting off 
the tails of the data), and obtained the following results. $j=10^2, \; \lambda = 0.54 \; (0.34)$;
$j=10^3, \; \lambda = 0.49 \; (0.39)$;  $j=10^4, \; \lambda = 0.58 \; (0.44)$; 
$j=10^5, \; \lambda = 0.54 \; (0.35)$; $j=10^6, \; \lambda = 0.62 \; (0.47)$; and
$j=10^9, \; \lambda = 0.63 \; (0.39)$. The figures in parentheses stem from
Fig.~3. There is a rough agreement, but certainly 
also   considerable scatter in these data, with the log-wait time 
type of analysis yielding systematically lower figures.  At this stage
it is unclear whether the  non-monotonic dependence of $\lambda$ 
on $j$ seen in Fig.~3
 is a real effect - or just due to a combination of statistical fluctuations
and systematic deviations from the ideal log-Poisson behavior.   
 
Summarizing the results from Figs.~2 and 3,
it appears that the value of ${\bf K}$ very strongly affects the average slope of the
curves (i.e. the value of $\lambda$). 
 The  degree of terracing 
might also have  an  effect on the slope, albeit a minor one.  

The independence of the different $\Delta_k$'s 
implied by the record statistics 
was tested by calculating the 
correlation coefficients $C_{ij}$ between $\Delta_i$ and $\Delta_j$. In practice, we
checked for $C_{12}$, $C_{23}$ and $C_{13}$. As   expected, the highest degree
of correlation was found for the relatively smooth landscape
 with ${\bf K} =7$.
In this case, the $C$ values were close to $0.4$.
 For ${\bf K} =31$ the correlation  
coefficients were of the order of $0.1$, i.e. of the same order as the 
statistical sampling error. 
\begin{figure}[t]
\centering
\includegraphics[width=0.7\textwidth]{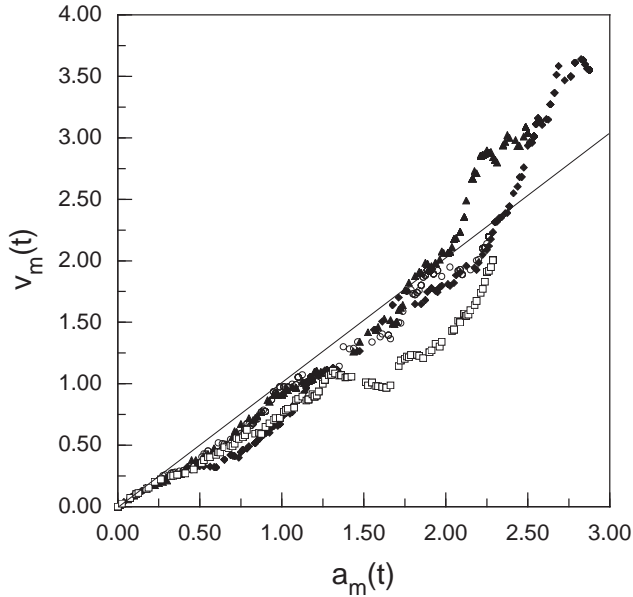}
\vspace{-3cm}
\caption{Sample variance, $v_m$, of the number of fitness jumps  versus   sample average, $a_m$,
of the same quantity. The data are all for ${\bf K}=31$.  Squares and filled diamonds are both for
for unterraced landscapes ($j=10^9$), and have mutation rates $u=10^{-4}$ 
and $u=5 \cdot 10^{-4}$ respectively. The other two data sets  both have $u=5 \cdot 10^{-4}$.
The degree of terracing is $j=10^2$ (circles) and $j=10^3$ (filled triangles). }
\label{var_vs_ave}
\end{figure}

To conclude the description of our data  
we plot in Fig.~4 the estimated variance $v_m(t)$ versus
the estimated average $a_m$, for ${\bf K}=31$ and a number of terrace values.
For a perfect agreement with the log-Poisson distribution
 the points should lie on a straight line of slope $1$. This is 
close to the observed behavior,  except for the highest values (where, on the other 
hand,  the statistics is poorest). A similar plot for ${\bf K}=15$ and ${\bf K}=7$
shows a systematic deviation from a straight line, with considerably less
variance than  in a purely random case.
Additional details on the simulations and on the genetic algorithm
utilized to produce them can be found in Ref.~\cite{Pedersen99}.
 
In summary, the dynamics of our evolutionary model
is  time inhomogeneous stochastic  process, 
with a rate of events falling off as $1/t$. The  
log-Poisson statistics   describes the data best  
for landscapes  with   large  $\bf K$, with or without
terraces.  In this  respect terraces have a minor
effect on the dynamics.

\noindent {\bf Discussion.}
While the shape of  the fossil record   certainly reflects   many
different factors\cite{Newman99b}, including e.g.  biogeography\cite{Pelletier99}
and external perturbations of the abiotic environment\cite{Alvarez80}, 
 the    event statistics demonstrated here   
should  have   rather  general  implications for  
all models  which do not completely dismiss the influence of population
dynamics on macroevolution.
If the evolutionary `jumps'  are the elementary events
in any dynamics,  possibly involving interactions
between evolving species, using $\log t$ as the
independent variable makes this dynamics appear as a stationary,
markovian process,  described e.g. by a master  or 
Fokker-Planck equation. 
The eigenvalues (and eigenvectors) of the evolution 
equation  describe the relaxation of any  average of interest. 
Since   the  observational time window is usually narrow 
in  $\log$ time, one is  restricted to
observing the decay of a  single (or a few) relaxational mode(s). 
As an exponential function of   $\log$ is a power-law,
the above  mechanism  offers a generic explanation 
for the power-law like behavior  found in several evolutionary
patterns.

\noindent {\bf Aknowledgments.} Both authors 
would like to thank Mark  Newman and   Richard Palmer
for inspiring conversations and exchanges of
ideas at the Santa Fe Institute  of Complex Studies
and at the Telluride Summer Research Institute.
Parts of this work  were commenced during 
a visit by A. P. at Duke University.  
A.  P. owes a special thank 
to Richard Palmer for the  kind hospitality
extended to him  and for the 
guidance  he  received during his stay. This project was partly supported by a block 
grant from   Statens Naturvidenskabelige Forskningsr\aa d.


\begin{thebibliography}{10}

\bibitem{Wright31}
S. Wright
\newblock{\it  Genetics} {\bf  16}, 97 (1931) 

\bibitem{Holland75}
J. H. Holland
\newblock{\it  Adaptation in Natural and Artificial Systems}. U. of Michigan Press. (1975)

\bibitem{vanNimwegen97}
E. van Nimwegen, J. P. Crutchfield and M. Mitchell
\newblock {\it Phys. Lett.}, {\bf A 229}, 144 (1997)

\bibitem{Eldredge72}
N. Eldredge and S. J. Gould {\em  Punctuated equilibria: An alternative to phyletic gradualism}
Models in Paleobiology. (T. J. M. Schopf, ed.). Freeman, Cooper, San Francisco (1972) 


\bibitem{Lenski94}
Richard E. Lenski and Michael Travisano
\newblock {\it Proc. Natl. Acad. Sci. USA}, {\bf 91}, 6808  (1994), and
Dynamics of adaptation and diversification, in
 {\it Tempo and mode in evolution}, Walter M. Fitch and
Francisco J. Ayala Ed., National Academy of Sciences  (1995) 

\bibitem{Kauffman95}
Stuart Kauffman
\newblock {\it At home in the universe}, Chapt. 9, Oxford University Press (1995) 

\bibitem{Aranson97}
I. Aranson, L. Tsimring and V. Vinokur
\newblock {\it Phys. Rev. Lett.}, {\bf 79}, 3298 (1997)

\bibitem{Raup82}
David M. Raup and J. J. Sepkoski, Jr.
\newblock {\it Science}, {\bf 215}, 1501 (1982)

\bibitem{Newman98}
M. E. J. Newman and Gunther J. Eble,
\newblock {\it   Paleobiology}, {\bf 25}, 434, (1999)  

\bibitem{Sibani95}
Paolo Sibani, Michel R. Schmidt and Preben Alstr\o m
\newblock {\it Phys. Rev. Lett.}, {\bf 75}, 2055 (1995)

\bibitem{Sibani97}
Paolo Sibani
\newblock {\it Phys. Rev. Lett.}, {\bf 79}, 1413 (1997)

\bibitem{Alstrom99}
Preben Alstr\o m
\newblock {\it europhysics news}, {\bf 30}, 22 (1999)
 
\bibitem{Alvarez80}
Luis  W. Alvarez, Walter Alvarez, Frank  Asaro and Helen  V. Michel
\newblock{\it Science} {\bf  208}, 1095 (1980)

\bibitem{Raup84}
D. M. Raup and J. J. Sepkoski, Jr.
\newblock{\it Proc. Natl. Acad. Sci. USA} {\bf  81}, 801 (1984) 

\bibitem{Newman99b}
M. E. J. Newman and R. G. Palmer
\newblock {\it www.santafe.edu/$_{\tilde{\hspace*{0.1cm}}}$mark/pubs.html}
Models of extinction: a review.  
See also: Paolo Sibani, Michael Brandt and Preben Alstr\o m
\newblock {\it Int. J. Mod. Phys.}, {\bf B 12}, 361, (1998) 

\bibitem{Kimura83}
M. Kimura. {\em The neutral theory of molecular evolution}
Cambridge University Press 1983.

\bibitem{Bornholt98}
Stefan Bornholt and Kim Sneppen
\newblock {\it Phys. Rev. Lett.}, {\bf 81}, 236 (1998) 

\bibitem{Kauffman87}
S. A. Kauffman and S. Levine
\newblock {\it J. Theor. Biol.}, {\bf 128}, 11 (1987)  

\bibitem{Weisbuch92}
 G. Weisbuch, in {\em Spin glasses and biology,}
 edited by D.~L.~Stein (World Scientific, Singapore, 1992), pp. 141--158 

\bibitem{Newman98b}
M. E. J. Newman and Robin Engelhardt
\newblock {\it  Proc. R. Soc. London}, {\bf B 265}, 1333,  (1998) 

\bibitem{Sibani93}
Paolo Sibani and Peter B. Littlewood
\newblock {\it Phys. Rev. Lett.}, {\bf 71}, 1485 (1993)

\bibitem{Pedersen99} 
Andreas Pedersen  
{\em Master Thesis: Evolutionary Dynamics}, Odense Universitet, (1999),
see: \newblock {\it http://planck.fys.ou.dk/$_{\tilde{\hspace*{0.1cm}}}$ap/Public/thesis.ps}
 
\bibitem{Pelletier99}
Jon D. Pelletier
\newblock {\it Phys. Rev. Lett.}, {\bf 82}, 1983 (1999)
  
\end{thebibliography}
\end{document}